\documentclass[sigconf]{acmart}

\copyrightyear{2025}
\acmYear{2025}
\setcopyright{acmlicensed}\acmConference[CCS '25]{Proceedings of the 2025 ACM SIGSAC Conference on Computer and Communications Security}{October 13--17, 2025}{Taipei, Taiwan}
\acmBooktitle{Proceedings of the 2025 ACM SIGSAC Conference on Computer and Communications Security (CCS '25), October 13--17, 2025, Taipei, Taiwan}
\acmDOI{10.1145/3719027.3765027}
\acmISBN{979-8-4007-1525-9/2025/10}

\def\preprintversion{1}

\usepackage{xspace}
\usepackage{xcolor}

\usepackage{multicol}
\usepackage{multirow}

\usepackage{pifont}
\usepackage{adjustbox}
\usepackage{minted}
\usepackage{tabularx}

\definecolor{sangria}{rgb}{0.57, 0.0, 0.04}
\definecolor{pinegreen}{rgb}{0.0, 0.47, 0.44}
\definecolor{rossocorsa}{rgb}{0.83, 0.0, 0.0}
\definecolor{ao}{rgb}{0.0, 0.0, 1.0}
\definecolor{deepjunglegreen}{rgb}{0.0, 0.29, 0.29}
\definecolor{dartmouthgreen}{rgb}{0.05, 0.5, 0.06}

\newcommand{\eg}{e.g.,\xspace}
\newcommand{\ie}{i.e.,\xspace}
\newcommand{\etal}{et~al.\@\xspace}
\newcommand{\cf}{cf.\@\xspace}

\newcommand{\textttWithBreak}[1]{\mintinline[breaklines, breakafter={\_=-}]{text}{#1}}

\newcommand{\toolname}{\textsc{SbxBrk}\xspace}%

\newcommand{\bugreport}[1]{\href{https://issues.chromium.org/u/2/issues/#1}{\color{black}{\##1}}}
\newcommand{\bugreportlong}[1]{\href{https://issues.chromium.org/u/2/issues/#1}{\color{black}{\##1}}}
\newcommand{\blackurl}[2]{\href{#1}{\color{black}{#2}}}

\newcommand{\numbugs}{19\xspace}

\newcommand{\asan}{ASan\xspace}
\newcommand{\wasm}{Wasm\xspace}
\newcommand{\webassembly}{Web\-Assembly\xspace}
\newcommand{\javascript}{Java\-Script\xspace}

\def\v8{V8\xspace}
\newcommand{\rlbox}{RLBox\xspace}

\newcommand{\aflpp}{\textsc{AFL++}\xspace}

\newcommand{\fuzzilli}{\textsc{Fuzzilli}\xspace}
\newcommand{\fuzzillisbx}{\textsc{FuzzilliSbx}\xspace}
\newcommand{\libafl}{\textsc{LibAFL}\xspace}

\newcommand{\Ignition}{\textsc{Ignition}\xspace}
\newcommand{\Sparkplug}{\textsc{Sparkplug}\xspace}
\newcommand{\TurboFan}{\textsc{Turbofan}\xspace}
\newcommand{\turboshaft}{\textsc{Turboshaft}\xspace}
\newcommand{\Maglev}{\textsc{Maglev}\xspace}

\newcommand{\Torque}{\textsc{Torque}\xspace}
\newcommand{\CodeStubAssembler}{\textsc{Code Stub Assembler}\xspace}
\def\CPP{{C\nolinebreak[4]\hspace{-.05em}\raisebox{.35ex}{\footnotesize\bf ++}}\xspace}

\newcommand{\circleone}{\ding{202}\xspace}
\newcommand{\circletwo}{\ding{203}\xspace}
\newcommand{\circlethree}{\ding{204}\xspace}
\newcommand{\circlefour}{\ding{205}\xspace}
\newcommand{\circlefive}{\ding{206}\xspace}

\begin{document}

\title{Empirical Security Analysis of Software-based Fault Isolation through Controlled Fault Injection}

\author{Nils Bars}
\affiliation{
  \institution{CISPA Helmholtz Center for Information Security}
  \city{Saarbruecken}
  \country{Germany}
}
\email{nils.bars@cispa.de}

\author{Lukas Bernhard}
\affiliation{
  \institution{Independent}
  \city{Dortmund}
  \country{Germany}
}
\email{lukas.bernhard@rub.de}

\author{Moritz Schloegel}
\affiliation{
  \institution{Arizona State University}
  \city{Tempe}
  \state{AZ}
  \country{USA}
}
\email{moritz.schloegel@asu.edu}

\author{Thorsten Holz}
\affiliation{
  \institution{Max Planck Institute for Security and Privacy}
  \city{Bochum}
  \country{Germany}
}
\email{thorsten.holz@mpi-sp.org}

\begin{abstract}
We use browsers daily to access all sorts of information. %
Because browsers routinely process scripts, media, and executable code from unknown sources, they form a critical security boundary between users and adversaries.
A common attack vector is \javascript, which powers complex web interactions but exposes a large attack surface due to the sheer complexity of modern \javascript engines.
To mitigate these threats, modern engines increasingly adopt software-based fault isolation (SFI).
A prominent example is Google's \emph{\v8 heap sandbox}, which represents the most widely deployed SFI mechanism, protecting billions of users across all Chromium-based browsers and countless applications built on Node.js %
and Electron.
The heap sandbox splits the address space into two parts: one part containing trusted, security-sensitive metadata, and a sandboxed heap containing memory accessible to untrusted code. 
On a technical level, the sandbox enforces isolation by removing raw pointers and using translation tables to resolve references to trusted objects. Consequently, an attacker cannot corrupt trusted data even with full control of the sandboxed data, unless there is a bug in how code handles data from the sandboxed heap.
Despite their widespread use, such SFI mechanisms have seen surprisingly little security testing. %

In this work, we propose a new testing technique that faithfully models the security boundary of modern SFI implementations. %
Following the SFI threat model, we assume a powerful attacker who fully controls the sandbox’s memory. We implement this by instrumenting memory loads originating in the trusted domain and accessing untrusted, attacker-controlled sandbox memory.
We then inject faults into the loaded data, aiming to trigger memory corruption in the \emph{trusted domain} that processes this untrusted input. 
We implement our approach in a tool called \toolname and evaluate it on the \v8 heap sandbox.
In a comprehensive evaluation, we identify \numbugs security bugs in \v8 that enable an attacker to bypass the sandbox. 

\end{abstract}



\ifnum\preprintversion=0
\settopmatter{printacmref=true, printccs=true, printfolios=false} %
\else
\setcopyright{none} %
\settopmatter{printacmref=false, printfolios=true,printccs=true} %
\renewcommand\footnotetextcopyrightpermission[1]{} %
\pagestyle{plain}
\fi

\maketitle

\section{Introduction}
Web browsers are the primary way to access the Internet, acting as gateways to information, communication, and online services. This central role also makes them highly attractive targets: Browsers routinely process untrusted web content, including scripts and executable code.
A security vulnerability in the browser, such as a memory corruption bug, logic flaw, or improper isolation of web content, can lead to severe consequences, including data breaches, malware infections, and full system compromise.
In particular, executing attacker-supplied code, such as \javascript or \webassembly, poses a significant security risk, since this extends the attack surface from processing untrusted \emph{data} to executing untrusted \emph{code}.
Making matters worse, modern \javascript engines rely on complex mechanisms like \emph{Just-In-Time (JIT)} compilation and \emph{On-Stack Replacement (OSR)}~\cite{aycock2003brief,d2018stack}, which dynamically generate optimized machine code at runtime but also introduce vulnerabilities, for instance by removing runtime checks based on flawed assumptions. This complexity creates a large and complex attack surface that is difficult to secure.

To address this issue and confine untrusted code (called \emph{confinement problem}~\cite{lampson1973note}), systems employ a technique called \emph{Software-based Fault Isolation} (SFI)~\cite{wahbe1993efficient,morrisett2012rocksalt,erlingsson2006xfi, 234966, 255298, 7163016}.
This technique has been widely used in various security-sensitive applications~\cite{nacl,v8heapsbxblockpost,narayan2020retrofitting}. 
SFI confines the execution of potentially malicious or buggy code to a designated \emph{fault domain}, ensuring its effects remain isolated from the rest of the system. 
SFI restricts untrusted code to a specific memory region and defines a strict interface for interacting with the rest of the system. This interface defines the security boundary and is typically enforced through runtime checks. %
To protect web browsers, Narayan~\etal proposed an SFI scheme called \rlbox~\cite{narayan2020retrofitting}. This framework is used by the Mozilla Firefox browser to isolate third-party libraries by executing them in a \webassembly sandbox.
Firefox has used \rlbox to sandbox libraries, such as font~\cite{rlboxLibgraphiteImpl}
and XML parsers~\cite{rlboxLibxpaImpl}, significantly reducing the impact of vulnerabilities in these components~\cite{rlboxLibxpaBug1Impl}.
However, unlike such libraries, \javascript execution engines are tightly coupled to the browser and can hardly be split into atomic components that can be put into disjoint fault domains.

To tackle this problem, Google introduced the \emph{heap sandbox} SFI-mechanism in their \javascript and \webassembly engine \v8~\cite{v8designdoc}.
Given that \v8 is not only used by all Chromium-based web browsers but also by Node.js, Electron, Cloudflare's Workerd, and many other systems, this heap sandbox is arguably the most widely deployed SFI implementation in practice, protecting billions of users.
The development of the heap sandbox has been driven in particular by the observation that many security bugs found in the \v8 engine started as logic bugs in the JIT compiler~\cite{v8exploitablebug1, v8exploitablebug2, v8designdoc}.
Importantly, this problem cannot be mitigated by using memory-safe languages like Rust or modern hardware-backed mitigations, such as \emph{Memory Tagging Extension} (MTE)~\cite{Armv8.5MTE,serebryany2018memorytagging} and \emph{Control-Flow Integrity} (CFI)~\cite{abadi2009control,burow2017control}. While such techniques help catch or prevent many classes of memory safety issues in general, they do not address vulnerabilities introduced by \emph{logic errors} in JIT-compiled code.
Hence, Google opted to implement an SFI scheme, the heap sandbox, to protect users even when logic bugs allow the attacker to compromise the \javascript engine's heap.
Based on the assumption that logic bugs are inherent to browsers and that objects on the heap eventually get corrupted, the memory is split into two fault domains.
The first domain is the untrusted heap sandbox, which contains most heap objects accessible by the executed \javascript and \webassembly code, and therefore may be corrupted anytime.
The second domain is the trusted domain. It contains all security-critical data structures that must be protected from tampering, such as \CPP stack and heap, control flow metadata, memory management structures, and JIT-compiled code.
To prevent an attacker from leveraging corrupted objects in the sandbox to escape into the trusted domain, the data structures in the untrusted region are redesigned to remove unsafe primitives such as raw pointers or large offsets. Instead, offsets relative to the sandbox base express all references within the heap sandbox. If a reference to the trusted domain is needed, it is resolved through a protected lookup table.
This design ensures that code that processes malformed sandbox objects cannot be used as a vehicle to break out of the sandbox, because the necessary primitives for arbitrary memory access are absent.
While the lack of raw pointers makes it harder for attackers to directly corrupt trusted memory, bugs can still occur. For instance, if a value from the untrusted domain is used without sanitization as an index into a stack-allocated array, it may still lead to corruption outside the sandbox.
Consequently, it is crucial that the engine treats \emph{all} data read from it, \ie data crossing the \emph{fault domain boundary}, with utmost care. Rigorous security testing is needed to ensure all such code is free of bugs that would allow an attacker to escape the sandbox. Naturally, we expect this to include \emph{fuzzing}, a dynamic testing technique that has proven immensely effective in identifying bugs~\cite{scharnowski2022fuzzware, aschermann2019redqueen,bernhard2024darthshader}, including those in \javascript engines~\cite{groß2023fuzzilli, bernhard2022jit-picker}.

Unfortunately, all existing fuzzing techniques are unsuited for testing the fault domain boundary of \v8's heap sandbox.
Their strength lies in exposing memory safety issues in the \textit{front-end}, \ie the part of a system responsible for parsing and processing untrusted inputs.
However, this focus on the front-end fails to test the security guarantees provided by SFI approaches, as they assume a different threat model: The attacker has already exploited a vulnerability and now has arbitrary memory access, but only within a restricted execution environment. The goal is to escape that environment and compromise the trusted domain.
Consequently, a more sophisticated approach is needed.

In this paper, we present a novel fuzzing method for testing SFI implementations. We inject faults into every memory load where trusted code reads data from the untrusted domain.
This approach models an attacker with complete control over the contents of the untrusted heap memory who attempts to escape the sandbox to elevate privileges.
In the first step, we identify all points where data might pass the trust boundary from the untrusted heap sandbox to the trusted domain. 
Next, we filter out \emph{safe} accesses where load instructions are guaranteed not to access data inside the heap sandbox.
For the remaining loads, we employ inline runtime checks to determine if untrusted memory is accessed.
We then inject faults by mutating values crossing the boundary, simulating adversarial input from a compromised sandbox. As a feedback loop, we use code coverage to identify faults that trigger new execution paths in the target.
Overall, this approach enables us to reveal vulnerabilities in how the trust boundary is enforced by systematically corrupting data as it moves from the heap sandbox into trusted code.
We implemented our approach in a prototype tool called \toolname and evaluated it on the \v8 heap sandbox. Our fault injection-based method successfully uncovered bugs that break \v8's SFI guarantees and compromise the sandbox.
We found \numbugs security bugs in the \v8 heap sandbox, including stack-based buffer overflows, use-after-free issues, and double-fetch vulnerabilities that evaded the heap sandbox.
Our results show that even widely deployed SFI mechanisms can contain serious security-relevant bugs, underscoring the need for more effective testing methods to ensure the secure isolation of untrusted code.

\smallskip \noindent
\textbf{Contributions.} In summary, we make the following contributions:

\begin{itemize}
\item We propose the first automated testing approach that is aware of and tailored to software-based fault isolation.\vspace{5pt}
\item We implement our approach in a tool called \toolname and focus on Google Chrome's \v8 \javascript engine, which represents the most important software-based fault isolation mechanism in practice.\vspace{5pt}
\item We uncover \numbugs security bugs in \v8, underlining the practical relevance of our work.
\end{itemize}

\paragraph{Ethical Considerations and Open Science}
All vulnerabilities found have been responsibly disclosed to the Google Chrome team. Per Google's policy, details were withheld for (up to) 90 days and then made public. To foster future research on this important topic, we release our source code at \url{https://github.com/SbxBrk}. %

\section{Technical Background}%
\label{sec:background}
First, we provide a brief overview of the relevant background knowledge, including software-based fault isolation, the architecture of Google's \v8 engine, and the technical details of \v8's SFI mechanism, the heap sandbox.

\subsection{Software-based Fault Isolation}
Various fault isolation techniques have been developed to contain unintended or possibly malicious behavior of applications. 
A well-known approach is \emph{process-based isolation}, where each application runs in its own virtual address space, a standard mechanism supported by all major operating systems~\cite{denning1970virtual, Huck1993ArchitecturalSF}. This ensures that any fault, such as a crash or a vulnerability that allows arbitrary memory access, is contained within the boundaries of the faulty application's memory space. As a result, such faults do not directly impact other software running on the system. Unfortunately, such strong isolation comes with a significant performance and memory overhead. 
For instance, when a program uses an untrusted shared library and wants to isolate it, placing it in a separate process requires a costly inter-process communication mechanism to be employed.
This reduces efficiency and increases the system's complexity, potentially introducing new bugs~\cite{v8mojobug1, v8mojobug2, v8mojobug3, v8mojobug4}.

To mitigate the performance impact while still enforcing isolation, \emph{software-based fault isolation} (SFI) has emerged~\cite{wahbe1993efficient,morrisett2012rocksalt,erlingsson2006xfi,nacl,mccamant2005efficient}.
SFI assigns each component (\eg a shared library or plugin) its own \emph{fault domain}, a logically separate portion of the application's (virtual) address space~\cite{wahbe1993efficient}. 
To enforce the new security boundary between the \textit{fault domains}, code is rewritten or instrumented to ensure that all memory accesses and control-flow transfers (\eg jumps or calls) stay within their designated \textit{fault domain}.
This confines each component and prevents faults from spreading across the newly established \textit{fault domain boundary}.

In practice, \webassembly (\wasm)~\cite{WebAssemblyCoreSpecification} is one of the most widely used SFI implementations. 
\webassembly executes code inside a sandboxed environment, isolating it from the embedding application and other components. It enforces a structured memory model that prevents arbitrary memory accesses, ensuring that each code module operates within its own sandboxed environment.
This is achieved through techniques such as \emph{linear memory addressing}, which restricts memory access to a linear memory region, and \emph{bytecode validation}, which ensures that only safe instructions are executed.
Building on the isolation provided by \webassembly, \rlbox~\cite{narayan2020retrofitting} enables safe execution of code written in memory-unsafe languages within the same process. 
Mozilla uses \rlbox to compartmentalize potentially erroneous components, such as third-party libraries for parsing fonts~\cite{rlboxLibgraphiteImpl} or decoding audio content~\cite{rlboxLiboggImpl}. 
By compiling the unsafe code to \webassembly, \rlbox ensures that faults are contained and that memory safety and control-flow integrity are maintained outside the sandbox. 
Additionally, \rlbox employs a static type system to taint data to reduce the likelihood of these being used without proper sanitization.

\subsection{\v8 Engine}%
\label{sec:background:v8}
The \v8 engine is a highly complex engine that uses multiple tightly interleaved interpreters and compilers to facilitate efficient execution of \javascript and \webassembly code. 

\subsubsection{Code Generation}%
\label{sec:background:v8:codegeneration}
The \v8 engine employs both Ahead-Of-Time (AOT) and Just-In-Time (JIT) compilation techniques to execute \javascript and \webassembly efficiently. In the following, we explain these techniques in more detail.

\textbf{Ahead-Of-Time Compilation.}
AOT compilation occurs during the build process of the engine itself, where \v8's core runtime, including memory management, garbage collection, and built-in object representations, is compiled from \CPP using traditional compilers like Clang or GCC. 
This process also produces essential components such as the bytecode interpreter (\Ignition), the baseline and optimizing JIT compilers (\Sparkplug, \Maglev, \TurboFan, and \turboshaft), and support libraries, such as \Torque and the \CodeStubAssembler (CSA). 
\Torque, a domain-specific language (DSL), simplifies the implementation of built-in \javascript functions by allowing them to be written in a TypeScript-like syntax. 
These \Torque definitions are then transpiled into CSA, a lower-level abstraction that emits optimized machine code. 
In summary, to instrument the AOT-compiled code, the compilation of the \CPP code and the assembler used by CSA must be considered.

\textbf{Just-In-Time Compilation.}
Initially, \v8 compiles \javascript into bytecode, which the \Ignition interpreter executes using AOT compiled machine code stubs. 
To improve performance, the \Sparkplug JIT compiler translates this bytecode into machine code stubs that are linked using control flow instructions, thus eliminating the interpreter overhead of \Ignition. 
If the code needs to be optimized further, \Maglev provides better performance, while only moderately increasing overhead for compilation. 
For frequently executed functions, \TurboFan and \turboshaft, \v8's most advanced JIT compilers, recompile the code into highly optimized machine code using techniques such as inlining and speculative optimizations. 
If an optimization assumption is invalidated, deoptimization reverts execution to the lowest tier. 
All JIT tiers ultimately rely on \v8's raw assembler to emit machine code efficiently. 
For instrumenting JIT-compiled code, it is sufficient to instrument the machine code emitted by the assembler that is shared by all JIT tiers and also used by CSA for AOT compilation.

\subsubsection{Heap Sandbox}%
\label{sec:background:v8:sbx}
The \v8 heap sandbox~\cite{v8heapsbxblockpost} is a lightweight, in-process fault isolation mechanism designed for Chrome's \javascript and \webassembly execution engine \v8.
Its goal is to contain the effects of memory corruption originating from untrusted \javascript or \webassembly code and prevent it from being leveraged to escalate privileges by corrupting security-critical objects.

To enforce this boundary, the \v8 address space is split into two fault domains: a trusted domain, which holds security-sensitive objects, and an untrusted heap sandbox, which contains most heap objects accessible to \javascript and \webassembly code.
The data structures used in the untrusted portion of the engine must be adapted so that, even if they are corrupted, they cannot be leveraged to perform arbitrary memory reads or writes. This requires eliminating certain types, such as raw pointers or large integer offsets (\eg 64-bit offsets) that could otherwise be manipulated to reference arbitrary memory locations within the full process address space.
This transformation ensures that engine code processing attacker-controlled objects---essentially untrusted bytes---cannot inadvertently cause further memory corruption. The required primitives that could enable such propagation, such as raw pointers, have been completely removed from the untrusted domain.

However, removing raw pointers necessitates introducing a new way for objects to reference one another. This is done using offsets relative to a fixed base address rather than absolute pointers. That fixed base is the heap sandbox, a dedicated region of the address space reserved for untrusted heap objects. Limiting all relative references to this controlled region, the engine can enforce strict bounds on what corrupted objects can reference.

Access to trusted objects, such as JIT-compiled code or internal engine structures, is mediated through lookup tables. These tables act as controlled indirection layers, ensuring access is explicitly validated and cannot be redirected arbitrarily through corrupted data. Through this restructuring, \v8 ensures that even if memory corruption occurs within the sandbox, it cannot propagate beyond its boundary or affect the trusted parts of the engine.

On a technical level, the \v8 heap sandbox works by reserving a 1\,TiB virtual memory region in which most heap objects are confined. 
Within this cage, objects can reference other objects only in two controlled ways: either through a 40-bit offset relative to the base of the memory cage or via an index into a trusted translation table that resolves to the target's full 64-bit address. 
Consequently, an attacker cannot achieve memory corruption by simply overwriting pointers since all references inside the sandbox are either constrained to point to sandboxed data within the cage or are protected by trusted lookup tables that remain outside the attacker's control. 
Effectively, this confines arbitrary write and read primitives to the bounds of the sandbox.

To overcome this countermeasure, an attacker must find a bug in the sandbox's implementation that allows the corruption of data \emph{outside} the sandbox. 
Since the sandbox does not contain any raw pointers that can be leveraged for corruption, the goal is to find code \emph{outside} the sandbox that processes data originating from \emph{within} the sandboxed heap in an unsound fashion. 
A bug in such code may allow an attacker to achieve memory corruption that is not confined by the sandbox.
Unfortunately, this implies a vast attack surface that includes, for example, all built-ins mandated by the ECMAScript specification.
Such a built-in might allocate a stack buffer or a heap-allocated array (e.g., a \texttt{std::vector}). If data from the sandbox is copied into such buffers without proper size checks, an attacker can corrupt memory in the trusted domain.

\subsubsection{Attacker Model}%
\label{sec:background:attacker_model}
The attacker model used throughout this paper is defined by Google's guidelines~\cite{chromevrprules} on what is considered a security-relevant bug in \v8's heap sandbox. 
The attack model assumes that an attacker can execute arbitrary \javascript and \webassembly code and is equipped with arbitrary read and write primitives for the sandboxed heap memory. 
These primitives can also be used concurrently while code runs in a different thread, enabling the attacker to exploit TOCTOU and double-fetch issues (see Section~\ref{sec:background:corruption_api} for additional details).

The goal is then to break the isolation by corrupting memory \emph{outside} the sandbox boundary. To qualify as a security issue, a bug must demonstrate such an escape by causing memory corruption beyond the sandbox or, more severely, by performing a controlled memory write outside of it.
In short, the security model treats the sandbox as a barrier, and any ability to breach it through memory corruption is considered a significant vulnerability.

\subsubsection{Memory Corruption API}%
\label{sec:background:corruption_api}
To submit a bug reproducer to Google, the \v8 engine provides a \javascript API called \emph{Memory Corruption API}~\cite{v8heapsbxgitreadme} that provides the capabilities described in the attacker model above. In particular, this API allows for arbitrary locations of the heap sandbox to be modified.
\begin{listing}[tb]
\caption{Exemplary use of the Memory Corruption API to simulate heap corruptions within \v8.\vspace{-1em}} 
\label{lst:mem_corruption_api}
\begin{minted}[
frame=lines,
framesep=1mm,
fontsize=\footnotesize,
xleftmargin=0.2em
]{js}
// Get a view of the heap memory.
let memory = new DataView(new Sandbox.MemoryView(0, 0x100000000));
// Write into the heap memory
memory.setUint8(0xBBBBBBBB, 0x41)
\end{minted}
\end{listing}
For instance, the code displayed in Listing~\ref{lst:mem_corruption_api} would write the value \verb|0x41| to the address \verb|0xBBBBBBBB|. This API allows the implementation of proof-of-concepts (PoCs) and test cases that can be integrated into \v8's test suite. To perform concurrent memory modifications (to trigger TOCTOU and double-fetch bugs), \v8's Worker API~\cite{jswebworkerapi} can spawn background threads. These threads can then use the memory corruption API to perform the corruption concurrently.

\section{Design}%
\label{sec:design}
To effectively test the \v8 heap sandbox, arguably the most important real-world implementation of SFI, we need a new approach that is sensitive to the fine-grained boundary between the trusted and untrusted fault domain.

\subsection{Overview}%
\label{sec:design:overview}

\begin{figure}[tp]
    \centering
    \includegraphics[width=0.90\linewidth]{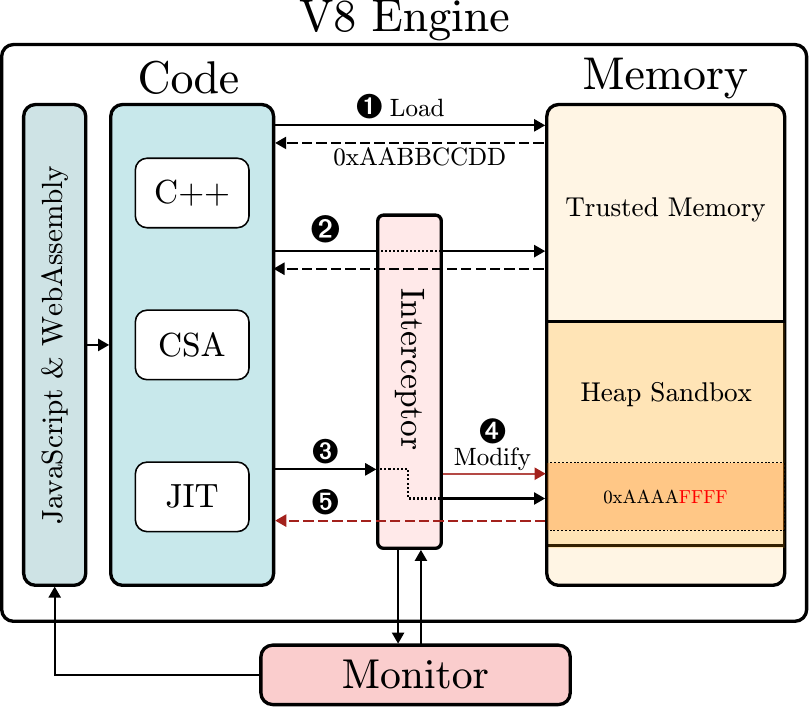}
    \caption{System overview of our prototype \toolname. All loads that may read from the untrusted \emph{Heap Sandbox} are instrumented, so they pass through the \emph{Interceptor}. The \emph{Monitor} controls whether a fault should be injected and is responsible for providing \javascript and \webassembly code as input.\vspace{-1em}}
    \label{fig:system_overview}
\end{figure}

Figure~\ref{fig:system_overview} displays an overview of our design and shows how the different components of the target, \v8, and our fuzzer interact.
The heap sandbox of \v8 separates the virtual address space of the \javascript and \wasm engine into two distinct fault domains.
The first domain is the \textit{heap sandbox} that is entirely untrusted and assumed to be fully controlled by an attacker (\cf Section~\ref{sec:background:attacker_model}). This memory region contains most of the engine's heap objects, \ie objects allocated by \javascript and \webassembly code. Corruptions in this memory region are contained in this fault domain because all references stored on this heap are relative to the sandbox's base, and pointers to external memory are resolved through lookup tables (see Section~\ref{sec:background:v8:sbx} for details).
The second fault domain is the \textit{trusted memory}. It contains all objects for which integrity must be maintained to ensure the security of the entire system.
Both fault domains are accessed by the code that constitutes the \v8 engine. Consequently, any code interacting with data in the heap sandbox is part of the attack surface. In particular, this means all of the components within \v8 that generate code must be considered to ensure we cover all interactions with the heap sandbox. As described in Section~\ref{sec:background:v8:codegeneration}, the code emitted by \v8's compilers and interpreters can be instrumented through two key techniques: modifying the assembler and adapting the \CPP code of the engine.
The assembler of \v8 must be extended to allow precise interception of memory accesses at the instruction level. Meanwhile, the \CPP code of \v8 can be instrumented using a compiler plugin that automates modifications at a higher abstraction level. Together, both modifications allow us to route all memory loads through the \textit{Interceptor} component.

From a testing point of view, memory loads can be grouped into the following three categories:
Loads of values where we can determine they are located in the trusted memory space (\circleone) at compile-time (of \v8). For example, global variables are stored in their designated memory region and will never be on the heap. Since our attack model (\cf Section~\ref{sec:background:attacker_model}) assumes that data in the trusted domain is benign, we can safely ignore these loads. We term such loads \emph{uninteresting}.
If a load never accesses the heap sandbox, we can avoid all runtime costs associated with instrumenting it. Thus, maximizing the number of loads in this category is desirable.

If we cannot determine a load as \emph{uninteresting} at compile-time, it is passed to a runtime component named \emph{Interceptor}. This component evaluates whether the load's destination is \emph{inside} or \emph{outside} the sandbox. Depending on the result, it is forwarded (\circletwo) or intercepted (\circlethree) if it targets data \emph{within} the heap sandbox. If we intercept a load, the \emph{Interceptor} can mutate the memory content before loading it (\circlefour). Such a mutation serves as our fault injection mechanism to test the code that handles data loaded from untrusted memory and will be discussed subsequently.
Eventually, the load is performed and returns a (mutated) value to the engine (\circlefive).

\subsection{Fault Domain Boundary}
All load instructions that access data in the untrusted heap sandbox constitute the \emph{fault domain boundary}. Our overarching goal is to inject faults into values that cross this boundary. 
Following the spirit of fuzzing, we posit that any such fault injection must be sound and highly efficient.
If it were unsound and we injected faults at locations that can never be under the attacker's control, the fuzzer would report false positive findings. 
Efficiency is crucial considering that the fault isolation mechanism isolates the heap, thus \emph{every} memory access potentially crosses the security boundary. As these accesses are abundant, our process must be highly efficient to allow for reasonable testing throughput.

\textbf{Naive Approach.} A naive approach to tackle the problem would be to simply intercept every memory read access in the program and check whether it crosses the boundary at runtime. This could be implemented by leveraging the memory management unit (MMU) to raise a page fault whenever memory within the heap sandbox is accessed.
However, this approach would introduce a significant slowdown for several reasons.
Both reads and writes must be intercepted, as MMUs typically do not support mapping pages as write-only. Consequently, interception would have to be enforced by marking pages as no-access, triggering faults on both read and write operations.
When a fault occurs, the fuzzer must temporarily grant access to the page to execute the faulting instruction. To maintain control afterwards, the fuzzer must either single-step through the instruction using a debugging interface like \verb|ptrace| or disassemble the instruction to insert a breakpoint just after it. Once the instruction finishes, the page must be marked as no-access again to catch future accesses. These steps impose a massive performance penalty (\cf Section~\ref{sec:eval:compare_with_simple_baseline}), making such a naive approach impractical for efficient fuzzing. 

\textbf{\toolname.} Instead, we propose the following approach:
\toolname leverages instruction-level instrumentation at compilation time, combined with a lightweight runtime component to efficiently identify and handle loads crossing the fault domain boundary.
As described in Section~\ref{sec:design:overview}, loads can originate from different compilers. More precisely, loads can be emitted either during the compilation of \CPP code and CSA, or during runtime by the JIT compiler. We discuss each of the respective origins and how we instrument them in the following.

\textbf{Loads Originating from \CPP.}
During compilation of \v8, we systematically analyze each load instruction to determine whether it is guaranteed to \emph{not} access the heap sandbox's memory and should not be considered for fault injection. We classify such load instructions as \emph{uninteresting}.
To identify the uninteresting instructions, we examine the destination of each load operation. Since the untrusted fault domain strictly encapsulates the heap, any load instruction that could be a valid target for fault injection must explicitly access heap memory. Consequently, we skip instructions loading from the stack and global variables, as these reside entirely within the trusted fault domain.
We consider a load relevant for fault injection when it is impossible to conclusively show that it is not tied to heap memory after traversing all control flow and data dependencies.
This procedure ensures that our approach remains sound and efficient. It filters out unnecessary fault injection targets early in the process while keeping all memory accesses that potentially cross the security boundary.
Eventually, we insert a call to our \emph{Interceptor} runtime before each \emph{interesting} load. We pass the memory address to be read and the width of the memory access, along with the call. Based on this information, the \emph{Interceptor} can decide at runtime whether the load targets the heap sandbox and whether faults should be injected before the memory is read.

\textbf{Loads Originating from CSA \& JIT.}
Some parts of the \v8 engine are implemented by leveraging CSA, which allows the specification of low-level code using \CPP. During the build process of \v8, a custom assembler lowers the code written using CSA into machine instructions. Fortunately, the same assembler is used when the JIT generates machine code during runtime.
As a result, extending the assembler once is sufficient to cover all remaining loads not emitted as part of the \CPP code. 

As with the \CPP code, we insert a call to the \emph{Interceptor} before each load instruction that potentially loads data from the heap sandbox. Because the instrumentation happens on a relatively low level, we do not have access to high-level information such as control or data flow. Consequently, we cannot easily prune loads that will never access untrusted memory. Theoretically, we could move the instrumentation to a higher abstraction layer (\ie the IR level used by the JIT) to filter loads, but this would make our approach considerably more complex. Consequently, we resort to conservatively instrumenting \emph{all} loads on this level and leave further optimization for future work.

\textbf{Interceptor.}
Finally, we propose to use a lightweight runtime component for all loads for which our static analysis cannot determine that they do not access heap memory.
This runtime check is invoked immediately \emph{before} each interesting load and ensures that the memory access falls within the bounds of the heap sandbox security domain. As we know the base address of the heap and its size, we can easily verify this. If the check determines an access is constrained within trusted memory, we do not consider this load for fault injection. On the other hand, if it accesses the heap sandbox contents, we can inject faults to test the security boundary.
This hybrid approach---combining static elimination with dynamic validation---minimizes our overhead while allowing sound instrumentation of the security boundary.

\subsection{Fault Injection}
Given this precise interception of all memory accesses crossing the fault domain boundary, we can now focus on data being passed from the untrusted domain to the trusted one without being properly sanitized.
We rely on \emph{software fault injection}~\cite{bars2023fuzztruction} to find such errors in trusted code handling untrusted data.
More precisely, the objective is to simulate an attacker who manipulates heap data just before it is accessed by trusted code. This approach allows us to uncover vulnerabilities that arise from incorrect assumptions about the untrusted data.

To trigger these memory loads---and thus enable fault injection---the \v8 engine must execute code that interacts with the sandbox. Our \textit{Monitor} component provides this input code in the form of \javascript files drawn from a static seed set. These files can also embed \webassembly bytecode to cover different areas of the engine. For simplicity, we refer to these inputs as \javascript throughout the remainder of this work, even if the \javascript embeds \webassembly. Importantly, these seeds are not generated or mutated by \toolname; instead, they must be produced up front by existing \javascript fuzzers, such as \fuzzilli~\cite{groß2023fuzzilli} or drawn from test cases included with \v8.

For each fuzzing iteration, the \emph{Monitor} provides a fuzzing input that is interpreted as a byte vector---this is the same input typically provided to a fuzzing target via a file or standard input. These bytes are then chunked into a sequence of fault injection bitmasks, with each \textit{interesting} load consuming one such mask in the order the loads occur.
Each bitmask is applied to the associated accessed memory location \emph{before} executing the actual load; in other words, we inject a (persistent) fault at the address from which the value will be loaded.
On a technical level, before executing an \emph{interesting} load, say $load(addr, 4)$ (reading four bytes from address $addr$), our instrumentation invokes the \emph{Interceptor}, passing address and size of the access ($interceptor(addr, 4)$).
The \emph{Interceptor} then (i) loads this value itself, $val = load(addr, 4)$, (ii) applies the bitmask $mod\_val = val \oplus bitmask$, and (iii) stores the modified value $store(mod\_val, addr, 4)$.
This effectively introduces a fault to the loaded value.
This way of fault injection yields two properties for the fault: \emph{localized} and \emph{persistent}.
Using a bitmask and $\oplus$ as the operator, most parts of the original value remain unchanged; \ie the fault is typically \emph{localized}. In line with the fuzzing spirit, our fault injection covers the full spectrum of modifications, ranging from flipping a single bit to altering the entire value. Most injected faults involve only minor changes. The intuition is that small modifications are more likely to trigger corner cases without violating the checks enforced by the code, whereas large changes often cause the input to be discarded.
At the same time, faults are also \emph{persistent}, as we store the modified value back in memory. Once a value is mutated, any later load from the same address receives the same faulty value. The fuzzer does not need to reapply or track the mutation across the program, which reduces complexity.

A crucial consideration is \emph{what} fuzzer input to provide to test deeper parts of the program effectively. First, we allow the bitmask to be zero, so the loaded value is not modified (since $v \oplus 0 = v$). This helps preserve the program’s normal behavior, allowing \toolname to explore more code paths that would otherwise break if the data were corrupted too early. The stream of fuzzer input initially consists of bytes containing only zeroes, \ie no mutations are applied. We use standard \aflpp-style (non-deterministic) mutation strategies, including mutation stacking, random bit flips, and appending additional bytes to the stream. These mutations are generic and not tailored specifically to our fault injection mechanism.

To determine whether a mutation of the fuzzer input exercised novel behavior, we use \emph{coverage feedback}. On a high level, there are three possible outcomes of our fault injection:
\begin{enumerate}
    \item It led to an early exit, as the program's checks considered the data to be malformed, 
    \item it led to \emph{no} observable difference in code coverage, meaning we have still explored the same code behavior, or
    \item it caused different behavior, resulting in new code coverage.
\end{enumerate} 
The first two outcomes are uninteresting for our purposes: Either the code checks the data safely, or our injected fault leads to no new behavior. In both cases, we do not want to keep this fault. However, when our injected fault leads to new code coverage, we can explore new program behavior. In this case, we store the fuzzing input (\ie bitmask) for further mutation, together with the \javascript seed file. Note that other than traditional fuzzing, \emph{we do not mutate the \javascript seed file}. Any set of \javascript files (some embedding \webassembly) can be used to execute \v8 and apply our fault injection approach to loads occurring at runtime. It is desirable, however, that this set covers as much functionality of \v8 as possible.

This utilization of \emph{coverage feedback} allows us to stick to a fuzzing-esque spirit and weed out uninteresting loads and mutations at runtime. This is due to the fuzzer's inherent novelty search, striving to find more program behavior and, thus, optimizing towards interesting injected faults.

\section{Implementation}%
\label{sec:implementation}

Our prototype \toolname is based on \libafl~\cite{fioraldi2022libafl}, and its core consists of roughly 5,000 lines of Rust code. For coverage instrumentation of \v8, we used the AFL++~\cite{AFLplusplus} compiler wrapper in version 4.22a.
As a compiler for building our instrumentation pass and as a backend for AFL++, we used LLVM commit \blackurl{https://github.com/llvm/llvm-project/commit/10c6d6349e51bb245b9deec4aafca9885971135b}{\texttt{10c6d6}}  of the LLVM 20 development branch.
To facilitate our approach, improve performance, and enhance the bug-finding capabilities, we applied patches to \v8 and LLVM. The patches and details regarding the implementation of our fuzzers are described below.

\textbf{LLVM Patches.}
While our approach does not strictly require modifications to LLVM, we customized our LLVM version by relaxing certain \asan checks. Specifically, we removed checks for out-of-bounds (OOB) memory \emph{reads} and for detecting overlapping memory regions passed as source and destination to \verb|memcpy| and similar functions. These checks were disabled because only memory corruptions \emph{outside} the heap sandbox, such as OOB \emph{writes} or use-after-free vulnerabilities, are considered security-relevant in our threat model (\cf Section~\ref{sec:background:attacker_model}). Discarding uninteresting bugs reduces noise and allows the fuzzer to discover more vulnerabilities that would otherwise be obscured by early termination due to an OOB read. We stress that this is not merely a theoretical concern, as demonstrated by the bugs we identified during our evaluation (see Section~\ref{sec:evaluation} for details) and prior work~\cite{raj2024fuzztothefuture}.

\textbf{Load Instrumentation.}
As described in Section~\ref{sec:design}, our instrumentation of load instructions must account for two sources of code: (1) the code generated by LLVM when compiling the \CPP code of \v8, and (2) the code generated by a custom assembler used during AOT compilation of \v8 and JIT compilation during execution.

For loads originating from the \CPP code, we use a custom LLVM pass that runs during \v8's compilation. This pass iterates over all load instructions and attempts to prune those that can be provably determined not to access the heap sandbox. It recursively tracks loaded values until it reaches either an \verb|AllocaInst| (stack allocation) or a \verb|GlobalVariable| (global allocation). For loads that (potentially) access the heap sandbox, the pass inserts an inline check directly before the load instruction. This avoids the overhead of saving the register state and calling the \emph{Interceptor} when the load does not touch sandboxed memory. If the check determines that the load targets the heap sandbox, control is transferred to \textttWithBreak{__fuzzer_before_heap_sandbox_load}, which receives the target address and the size of the load as arguments.

For loads emitted by the \v8 assembler, we instrument low-level operations, such as \verb|mov|, \verb|add|, and \verb|sub|, when their source operand references memory, \eg \verb|mov rax, [rbx]|. At each instrumentation site, we first save the \verb|FLAGS| register and some scratch registers before inserting a call to \textttWithBreak{__fuzzer_before_heap_sandbox_load_preserve}. This function follows the \textttWithBreak{preserve_all} calling convention and calls our interceptor. This intermediate step ensures that no registers are clobbered, which is vital since the newly added call instruction may be executed during arbitrary function execution.

\textbf{\v8 Engine Patches.}
We modified the \v8 engine to support the inline check described above. When the heap sandbox is allocated, a callback into our runtime is triggered. This callback sets the value of the \textttWithBreak{__fuzzer_heap_sandbox_base} variable, which holds the base address of the heap sandbox. This variable is exported by the runtime and used by various components, such as the LLVM pass, to quickly determine whether a load operation targets the heap sandbox.

Finally, we implemented a mechanism to pass \javascript files to \v8 via a memory buffer. This enables the \emph{Interceptor} runtime to efficiently supply \javascript input files without relying on traditional file system operations, thus reducing overhead.

\textbf{Interceptor Runtime.}
The \emph{Interceptor} runs within the \v8 target process and has two primary responsibilities: (1) receiving \javascript files and mutation byte masks via shared memory from the \emph{Monitor} and (2) forking the \v8 process and providing the newly created process with the fuzzing input received from the \emph{Monitor} component. During execution, it intercepts loads (using the \verb|__fuzzer_before_heap_sandbox_load| function called by our instrumentation) and potentially injects faults into memory based on the mutation byte masks. Eventually, the \emph{Interceptor} reports the exit status of the forked process to the \emph{Monitor}.

\textbf{Delayed Fork Server.}
To optimize \toolname, we implemented a fork server that can be dynamically placed at arbitrary points during execution of \v8. To mark all locations where the fork server can be placed, we introduced a new built-in \verb|FuzzerInjectionPoint()|, which can be inserted into seed \javascript files to indicate viable fork server positions.
This primitive is especially useful for skipping the initial startup phase of the \v8 engine, which involves setting up the \javascript execution environment and triggers significant interaction with the heap sandbox. This phase is not interesting for fuzzing because an attacker would never have control during it.
Additionally, bugs triggered during setup may prevent execution from reaching more diverse and test-case-specific code paths, ultimately limiting coverage.

\textbf{Monitor.}
The \emph{Monitor} is the central component of \toolname and is built on top of \libafl~\cite{fioraldi2022libafl}. It manages the fuzzing workflow by maintaining a queue of test cases that consist of \javascript files and the associated mutation byte masks. Test cases are passed to the \emph{Interceptor} via shared memory, which allows efficient communication.
During execution, the \emph{Monitor} tracks coverage information by reading data from the coverage map. This feedback guides the fuzzing process, helping to prioritize test cases that trigger new execution paths or uncover potential vulnerabilities.

\section{Evaluation}%
\label{sec:evaluation}
We now evaluate the performance of our prototype \toolname. 
The evaluation is split into three parts. (i) We evaluate \toolname's capabilities of finding bugs compared to a modified version of \fuzzilli~\cite{groß2023fuzzilli}. (ii) We perform several ablation studies to justify design decisions.
(iii) Last, we demonstrate the real-world impact of \toolname and discuss several security bugs it found.

\subsection{Setup}
In the following, we detail the setup used during evaluation. The design of the conducted experiments, and their documentation, follow the recommendations by Schloegel~\etal~\cite{schloegel2024sok} and by Klees~\etal~\cite{klees2018_evalfuzztesting}.

\textbf{Hardware Environment.}
We used the same hardware configuration for all our experiments: two AMD EPYC 9654 CPUs (totaling 192 physical cores / 384 logic cores), 768\,GB of RAM, and an SSD as backing storage. We used CPU pinning during the experiments to restrict fuzzer instances to one (logic) core each.

\textbf{Corpus.}
For experiments requiring a corpus of \javascript files that may embed \webassembly, we used \fuzzilli commit \blackurl{https://github.com/googleprojectzero/fuzzilli/commit/f31876fff984ef7050adcdbe7f4c8bb3255ed8d0}{\texttt{f31876f}} as a generator. The corpus was built on the hardware specified above. We launched 384 worker instances to maximize coverage and let them run for three days. Additionally, we passed the \verb|--wasm| flag to \fuzzilli, enabling it to generate \webassembly code inlined within the generated \javascript code.

\textbf{Target Preparation.}
For the preparation of \v8, we used, if not noted otherwise, the build options displayed in Table~\ref{tab:build_options} in the Appendix.
For compiling \v8, we used a different compiler depending on the fuzzer.
In case of \fuzzilli, we used the compiler bundled with \v8.
For \toolname, we relied on \aflpp's~\cite{AFLplusplus} compiler wrapper in version \blackurl{https://github.com/AFLplusplus/AFLplusplus/tree/75d8c47a6b8ae94cd7ded2f0574e4d35a2021ab7}{4.22a} by exporting \textttWithBreak{CC=afl-clang-fast} and \textttWithBreak{CXX=afl-clang-fast++}. Also, we set \textttWithBreak{-fpass-plugin=<path-to.so>} via \verb|CFLAGS| and \verb|CXXFLAGS| to enable our custom LLVM instrumentation pass. As the backing compiler for \aflpp, we used LLVM \blackurl{https://github.com/llvm/llvm-project/commit/3bd3e06f3fe418e24af65457877f40cee0544f9d}{\texttt{21.0.0}}.

For the \v8 engine itself, we used the following runtime flags: To disable certain functionality that is not fuzzing safe, \ie that may cause a false positive, we set \verb|--fuzzing|. Additionally, to ignore any error except heap sandbox corruptions, we use \verb|--sandbox-fuzzing|.
To increase the stability and determinism, the \verb|--single-threaded| flag was used, such that the engine does not spawn multiple threads on its own. Furthermore, to increase the attack surface, we also passed \verb|--allow-natives-syntax| and \textttWithBreak{--expose-gc}.

\subsection{Comparison with Baseline}%
\label{sec:eval:compare_with_simple_baseline}

\begin{figure}[tp]
    \centering
    \graphicspath{{images/}}
    \def\svgwidth{\linewidth}
    \begin{footnotesize}
        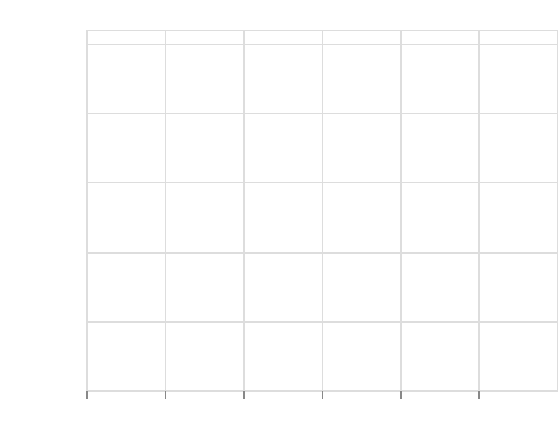
    \end{footnotesize}
    \vspace{-2em}
    \caption{The coverage (in \aflpp branches) \toolname and \fuzzillisbx achieve over ten 3-day runs on the \v8 engine. Displayed are the
median run, 66\% intervals, and the coverage achieved by the used seed corpus (\hspace{.6ex}\raisebox{.35ex}{\includegraphics[height=.45ex]{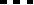}} ). \vspace{-1em}}
    \label{fig:fuzzilli_vs_sbxbrk}
\end{figure}

As we are the first to propose an approach sensitive to the security boundary introduced by software-based fault isolation, no other method is available to test \v8's heap sandbox. 
To evaluate the capabilities of our design, we implemented a baseline fuzzer by extending \fuzzilli, the state-of-the-art \javascript fuzzer maintained by Google~\cite{groß2023fuzzilli}. We call this modified version \fuzzillisbx. It integrates support for \v8's memory corruption API (see Section~\ref{sec:background:corruption_api}), allowing it to modify arbitrary memory locations of the heap sandbox.
Integrating this API into \fuzzilli turns it into a baseline that can fuzz the heap sandbox by generating calls to the memory corruption API. We will release this patch of \fuzzilli (around 350 lines) as part of our paper's artifact. 
Just like for the corpora generation, we based this custom fuzzer on the \fuzzilli commit \blackurl{https://github.com/googleprojectzero/fuzzilli/commit/f31876fff984ef7050adcdbe7f4c8bb3255ed8d0}{f31876f}. As the target version of \v8, we used commit \blackurl{https://github.com/v8/v8/commit/4715559d4fe2ce6e2c0f6de3c966347b6da6a489}{4715559} from December 22, 2024.

We stress that this extension of \fuzzilli is not a sophisticated fuzzer. However, due to the lack of other fault boundary-sensitive fuzzers, it can serve as a reasonable baseline to investigate whether the sandbox is simply untested, such that any fuzzer can find bugs, or if more sophisticated approaches like \toolname are needed.

For \fuzzillisbx and \toolname, we used the seed corpus described previously. Using this corpus, we executed both fuzzers for three days, ten times each. For \fuzzillisbx, we spawned 384 worker processes that synchronize their state when new inputs have been found. For \toolname, we executed 384 independent worker processes, simply because synchronization is currently not supported. %

To measure coverage, we replayed the inputs found by both tools using the \v8 binary \toolname used during fuzzing. We counted the number of edges covered using \aflpp's (collision-free) coverage map. Although best practice suggests using a dedicated coverage-instrumented binary~\cite{schloegel2024sok}, this was not feasible in our case: Recompiling \v8 with, for example, LLVM source code coverage influences the emitted load instructions, hence some may appear in a different order or not at all, making consistent replay unreliable.
To ensure fairness, we verified that our instrumentation did not expose additional coverage feedback unavailable to \fuzzillisbx. This guarantees that \toolname had no unfair advantage during fuzzing.

We display the coverage results of this comparative experiment in Figure~\ref{fig:fuzzilli_vs_sbxbrk}. Both fuzzers explore many new edges beyond those already covered by the seed corpus. While \fuzzillisbx has a slight edge over our approach in the very beginning, \toolname quickly catches up and surpasses \fuzzillisbx, achieving 16\% more coverage by the end of the three-day run. This indicates that our approach is more effective at triggering novel behavior, increasing the likelihood of uncovering bugs.

During this experiment, \fuzzillisbx found a single bug (see Table~\ref{tab:bugs}, \bugreport{388437270}) that triggered memory corruption outside the sandbox. 
In comparison, our prototype \toolname found six bugs, including the bug discovered by \fuzzillisbx. All six issues were reported to Google and have since been fixed. Table~\ref{tab:bugs} provides a summary of these bugs and is discussed in detail in Section~\ref{sec:eval:finding_new_vulns}.

\subsection{Ablation Studies}
We conducted five ablation studies to quantify the effects of our individual design decisions and better understand the overall performance of \toolname.

\newcounter{experimentCtr}
\stepcounter{experimentCtr}
\textbf{Experiment \arabic{experimentCtr}: Page Fault Performance.}
First, we ran an experiment to determine the lower bound of the performance impact of the naive approach (see Section~\ref{sec:design}). For this approach, we propose relying on page faults to intercept loads from the heap sandbox. This evaluation helps assess the feasibility of such a naive approach. To perform this assessment, we built \v8 with two different instrumentation methods:

\begin{enumerate}
\item \textbf{Software-based}: Each load is instrumented by inserting callbacks, as described in this paper.
\item \textbf{Hardware-based}: Instead of inserting a callback before a load, we insert an \verb|int3| instruction to simulate the overhead of a load interrupted by a page fault. Additionally, we installed a signal handler that calls into the \textit{Interceptor} as loads normally would.
\end{enumerate}

We created a corpus of 100 \javascript files by randomly selecting from the seed set used in Section~\ref{sec:eval:compare_with_simple_baseline}. 
We then executed the fuzzer for 30 minutes in both configurations, ten times each. To minimize nondeterminism and keep the corpus static, we disabled coverage feedback. We also disabled mutations because the mechanism used to simulate the hardware-based approach does not expose the target address of memory loads to our runtime. As a result, executions generally take longer, since inputs are less likely to terminate early due to injected faults, which ultimately reduces the number of executions observed.

Based on the collected performance data, we observed a significant slowdown of $2.8$x in terms of execution throughput for the hardware-based approach compared to the software-based version proposed in this paper. Specifically, the hardware-based configuration achieved an average of $9.2$ executions per second ($\sigma = 2.09$), while the software-based version reached $25.6$ executions per second ($\sigma = 2.08$).
We emphasize that this slowdown represents a lower bound and is likely higher in real-world scenarios for two reasons. First, in our test, only load instructions trigger interrupts. However, in a real scenario, page faults would also occur for store instructions, as CPUs typically do not support raising faults only for read accesses. 
Second, to maintain precise control over each load, the fuzzer must reliably step over the faulting instruction before resuming normal execution. This can be implemented in one of two ways: by setting a breakpoint immediately after the instruction or by using \texttt{ptrace}-based single-stepping.
Regardless of the method used, the mechanism involves several costly steps after each page fault:
(1) Temporarily adjusting page permissions to allow the faulting instruction to complete.
(2) Resuming execution.
(3) Catching the resulting signal or trap (breakpoint interrupt or \texttt{ptrace} event).
(4) Cleaning up: either removing the breakpoint or disabling single-stepping.
(5) Restoring original page protections.
(6) Resuming execution again.
These additional steps introduce a significant slowdown. Unfortunately, accurately measuring this impact would require developing a completely new fuzzer, which is beyond the scope of this ablation study. Instead, we conservatively report the overhead for the underapproximation of this naive approach and show that it already causes a significant slowdown (even though it represents a lower bound).

\stepcounter{experimentCtr}
\textbf{Experiment \arabic{experimentCtr}: Fuzzing without coverage feedback.}
For classical fuzzers, relying on code coverage proved to be an efficient way to improve the performance of the fuzzer. However, the approach presented in this paper is different from most other fuzzers. Therefore, whether it can benefit from using coverage as feedback is uncertain. To investigate this, we performed an ablation by disabling the coverage feedback for \toolname and otherwise performing the same experiment as described in Section~\ref{sec:eval:compare_with_simple_baseline}, running the fuzzer ten times for three days and comparing the number of bugs found afterwards.

During this experiment, the fuzzer without feedback could not uncover \emph{any} security-relevant bug. It identified only out-of-bounds reads and stack exhaustion bugs that were also found when using coverage. These bugs are irrelevant in our attacker model, and Google does not fix them (\eg \bugreport{384486462}). Thus, they are easily triggered.
This struggle to find security-relevant bugs indicates that coverage is crucial for the success of our technique.

\stepcounter{experimentCtr}
\textbf{Experiment \arabic{experimentCtr}: Load Pruning.}
We use static analysis during load instruction instrumentation to avoid instrumenting loads that are guaranteed not to access the heap sandbox. To measure the impact of this optimization, we ran an ablation study that compared two builds of \v8 (\verb|eeb2845|): one with load pruning enabled and one without.
When load pruning is disabled, all load instructions are instrumented. Each load is checked at runtime to see if it accesses the heap sandbox. With load pruning enabled, this check is omitted for loads that can be ruled out statically, avoiding the associated runtime cost.
To keep results consistent, we disabled coverage feedback and used a fixed seed corpus, as in the \textit{Page Fault Performance} experiment, reducing the nondeterminism throughout the execution. We executed each configuration for 30 minutes and repeated the experiment 10 times each.

Despite load pruning eliminating $52\%$ of the instrumented loads (from $1,430,287$ down to $742,225$), the performance gain was modest: Execution speed increased by only $3.53$ execs/s on average, from $70.62$ ($\sigma = 1.08$) to $74.15$ ($\sigma = 1.29$) execs/s. This equates to a performance increase of $5\%$.
This small gain suggests that our runtime check is already highly efficient.

\stepcounter{experimentCtr}
\textbf{Experiment \arabic{experimentCtr}: Compiler Effects on Loads.}
\label{sec:eval:abl:different_compiler_versions}
The heap sandbox security boundary is defined by the loads that access the sandboxed memory. While \CPP defines these loads on a high level, the compiler decides \emph{whether} and \emph{when} a load happens on the machine code level. This can lead to various problems affecting the heap sandbox's security. 
For example, depending on the compiler and the flags used, the compiler may choose to either temporarily keep a value on the stack and write it to the heap at the end of the function, or to store the value directly on the heap and load it back from the heap if needed later. In the former case, the function does not consume any attacker-controlled data, while in the latter case, the attacker can manipulate the on-heap data before it is read back.
Consequently, bugs may only get visible if the compiler is updated or the build flags are changed. 

Ideally, we could automatically determine if a change in the code emitted by the compiler introduces a new bug, but this is a challenging problem in practice. To understand the extent and impact of using different compiler versions, we use a proxy metric: the difference in the number of store and load instructions emitted per line of code for different compiler versions. 
We used LLVM versions \verb|10c6d63| (July 2024) and \verb|3bd3e06| (January 2025) to build \v8 (\verb|eeb2845|). For each of these builds, we used a custom LLVM pass that records a mapping of the source code location to the number of load instructions emitted. Based on these values, we can approximate the fraction of source code lines affected by different compiler versions. Notably, not all changes in the number of loads necessarily indicate a bug, and this approach is also incomplete, as the order and precise location of loads are crucial.

Across both builds, $260,434$ unique source code lines emitted at least one load. From these source code lines, $2.84\%$ changed the number of loads emitted if the LLVM compiler version is changed. This shows that the fault domain boundary is not static and that using two compiler versions only six months apart may change whether a bug can be triggered or not.
Our results in Section~\ref{sec:eval:finding_new_vulns} confirm that these effects are not just theoretical, as we discovered bugs that only surface due to differences in how the compiler lowers high-level \CPP into machine code.

\stepcounter{experimentCtr}
\textbf{Experiment \arabic{experimentCtr}: Delayed Fork Server.}
As part of our implementation, we introduced a technique that delays the fork server's start until after the target's initialization. The underlying hypothesis is that this improves fuzzing effectiveness by skipping the \v8 engine's start-up phase and skipping parts of the input \javascript file, allowing the engine to be fuzzed in a more targeted fashion. This matters because injected faults may crash the engine early, reducing the likelihood that later parts of the code are ever reached. By delaying the fork server, we increase the probability that code at the end of the input will be executed.

To evaluate the impact of this feature, we repeated the bug experiment described in Section~\ref{sec:eval:compare_with_simple_baseline}, but with the delayed fork server disabled.
Over three days, this configuration failed to uncover \emph{any} bug. Further manual analysis shows that the engine's initialization phase is the main bottleneck. During start-up, the \v8 engine creates many \javascript objects to expose the complete set of APIs mandated by the \javascript specification. Benchmarking this phase reveals that the fuzzer intercepts $193,639$ heap sandbox loads---consuming $737.89$\,KiB of the mutation mask---before the first line of \javascript input is executed. This overhead makes it much more likely for the fuzzer to get stuck during initialization, which our delayed fork server approach avoids.

\subsection{Finding New Vulnerabilities}%
\label{sec:eval:finding_new_vulns}

\begin{table*}[t]
    \centering
    \footnotesize
    \caption{All \numbugs security issues we found in \v8 during our evaluation. Bugs are either \textcolor{pinegreen!50!black}{fixed} by upstream or marked as \textcolor{gray!50!black}{duplicate}.\vspace{-0.7em}
    } 
    \label{tab:bugs}
    \begin{adjustbox}{max width=1.0\linewidth}
 \begin{tabularx}{\linewidth}{llX}
        \toprule
        \textbf{Bug ID} & \textbf{Status} & \textbf{Description}\\
        \midrule
        \bugreport{385775375} & \textcolor{pinegreen!50!black}{fixed} & \verb|Runtime_TypedArraySortFast| is susceptible to a double fetch attack that allows an attacker to execute \verb|std::sort| on an undersized buffer, resulting in out-of-bounds writes.  \\
        \bugreport{388193955}$^*$ & \textcolor{gray!50!black}{duplicate} & Passing an unusually large value to \verb|toExponential()| causes a stack-buffer-overflow.  \\
        \bugreport{388437270}$^*$ & \textcolor{pinegreen!50!black}{fixed} & Calling \verb|JSON.stringify| on a String value with a length of \texttt{0xffffffff} causes an integer overflow during serialization and, eventually, an out-of-bounds write.  \\
        \bugreport{388616182}$^*$ & \textcolor{pinegreen!50!black}{fixed} & The code path for converting a \texttt{Number} to a \texttt{String} with a particular radix via \verb|toString(16)| contains a double fetch bug. This is triggered by changing the \texttt{Number} to NaN after being checked not to be NaN, a value the subsequent code does not expect. \\
        \bugreport{389713719} & \textcolor{pinegreen!50!black}{fixed} & When constructing a \verb|MemoryChunk|, an on-heap pointer table index is used unsanitized to compute an address that is written to.  \\
        \bugreport{389970331}$^*$ & \textcolor{pinegreen!50!black}{fixed} & Converting a String of length \texttt{0xffffffff} via \verb|BigInt()| causes an integer overflow and a subsequent stack-buffer-overflow. \\
        \bugreport{390205877} & \textcolor{pinegreen!50!black}{fixed} & A variant of the \verb|CachedTieringDecision| \texttt{enum} is constructed from on-heap data that causes undefined behavior in the functions that use a non-exhaustive switch statement to compute their return value. \\
        \bugreport{390453039} & \textcolor{pinegreen!50!black}{fixed} & A variant of the \verb|AddressType| \texttt{enum} is constructed from on-heap data that causes undefined behavior in the functions that use a non-exhaustive switch statement to compute their return value. \\
        \bugreport{390568183} & \textcolor{pinegreen!50!black}{fixed} & A variant of the \verb|MessageTemplate| \texttt{enum} is constructed from on-heap data that causes undefined behavior in the functions that use a non-exhaustive switch statement to compute their return value. \\
        \bugreport{392541992} & \textcolor{pinegreen!50!black}{fixed} & Serializing a \verb|URIError| containing a \texttt{String} with a specific length via \verb|d8.serializer.serialize| causes a use-after-free bug.   \\
        \bugreport{392938085} & \textcolor{pinegreen!50!black}{fixed} & During parsing of the AST, one byte strings, \eg \verb|"A"|, are converted into \texttt{String} instances via a lookup table. This lookup process is subject to a double fetch bug that causes an out-of-bounds write. \\
        \bugreport{393989622} & \textcolor{pinegreen!50!black}{fixed} & Comparing two strings using \verb|localeCompare()| causes an out-of-bounds write for long strings. \\
        \bugreport{395029283}$^*$ & \textcolor{pinegreen!50!black}{fixed} & Converting a \texttt{Number} to a \texttt{String} via \verb|toPrecision| causes a buffer-overflow if a temporarily allocated \texttt{Number} object is mutated.   \\
        \bugreport{396446145}$^*$ & \textcolor{pinegreen!50!black}{fixed} & Inserting escape sequences into a \texttt{String} while it is processed by \verb|JSON.parse| causes a stack-buffer-overflow. \\
        \bugreport{397875195} & \textcolor{gray!50!black}{duplicate} & Construction of a \verb|FeedbackMetadata| \texttt{enum} variant causes an out-of-bounds write if converted to a \texttt{String}. \\
        \bugreport{398773898} & \textcolor{pinegreen!50!black}{fixed} & Calling \verb|JSON.stringify| on an unusually long property key may cause an out-of-bounds write. \\
        \bugreport{403372467} & \textcolor{pinegreen!50!black}{fixed} & Integer overflow in the icu library when formatting a malformed \verb|Number| causes an out-of-bounds write. \\
        \bugreport{411598604} & \textcolor{pinegreen!50!black}{fixed} & Use-after-free when performing tear down involving large pages. \\
        \bugreport{414831374} & \textcolor{pinegreen!50!black}{fixed} & An integer overflow in \verb|Module::GetModuleNamespace| causes an allocation to be smaller than expected, resulting in an out-of-bound write. \\
        \midrule
        \multicolumn{3}{l}{\footnotesize{}$^*$: Bug has been found as part of the experiments performed in Section~\ref{sec:eval:compare_with_simple_baseline}, \ie within three days.} \\
    \end{tabularx}
    \end{adjustbox}
    \vspace{-1em}
\end{table*}

To evaluate the capability of \toolname to find novel bugs, we performed a bug experiment by fuzzing \v8 for several days. During this campaign, we uncovered 10 bugs beyond those reported in Section~\ref{sec:eval:compare_with_simple_baseline}, totaling \numbugs bugs. These include heap out-of-bounds writes, stack-buffer overflows, and use-after-free vulnerabilities.
Each bug, including a description, is listed in Table~\ref{tab:bugs}. 
Besides the listed bugs, our fuzzer found multiple out-of-bound reads that caused a segmentation fault or stack overflow. However, according to Google (\eg \bugreportlong{384486462}), these are not considered relevant to security and are usually not fixed. We refrain from reporting such bugs to avoid overloading the developers with undesired reports. 
In the remainder of this section, we present several particularly interesting bugs in more detail and discuss what makes them unique to SFI.

\newcounter{caseStudyCtr}
\stepcounter{caseStudyCtr}
\textbf{Case Study \arabic{caseStudyCtr}: Double Fetch.}
One category of bugs \toolname found are \textit{double fetch} bugs that occur if a value is read multiple times from the heap sandbox, and the program implicitly assumes the value did not change between reads. 
Bugs that we found in this category include \bugreport{396446145}, \bugreport{392938085}, and \bugreport{385775375}.
While it is usually sufficient to treat values read from the heap sandbox as untrusted in a local context, these bugs require the developer to consider situations where constraints that held only a few lines before are unexpectedly obsolete.
Notably, exploiting such bugs requires attacker-controlled code to run \emph{concurrently} if the two loads are not naturally interleaved with code under the attacker's control. However, this condition is easily met under the assumed attack model (see Section~\ref{sec:background:attacker_model}).

We now take a closer look at one representative example of bugs of this type:
Bug \bugreport{385775375} is located in the \verb|TypedArraySortFast| function, which is used in multiple locations to sort typed arrays. The (pseudo) code containing the bug is shown in Listing~\ref{lst:bug_385775375_cpp}.
The problem here is that the size of the array, which is sorted, is fetched in two locations. First, the array length (in \emph{number of elements}) is stored in \verb|length| at the beginning of the function. Next, the content of the array is copied into a newly allocated buffer (1). This time, \verb|bytes| is used to determine the allocation size \emph{in bytes}.
Notably, this value might not match the one stored in the \verb|length| variable, as it was fetched from the heap again. Eventually, \verb|std::sort| is called (2), and the address of the last element is calculated using \verb|length|. Consequently, \verb|std::sort| can be tricked into sorting a larger array than allocated. Since the buffer is allocated on the trusted \CPP heap, this will lead to memory corruption outside the heap sandbox.

\begin{listing}[t]
\caption{Pseudo code containing bug \bugreport{385775375}. If the to-be-sorted buffer is shared (\texttt{copy\_data == true}), the sorted data is copied into a new buffer that is allocated in the trusted \CPP heap using \texttt{bytes} as size (1). After copying the data, it is sorted (2) using \texttt{length} to calculate the last element.\vspace{-1em}} 
\label{lst:bug_385775375_cpp}
\begin{minted}[
frame=lines,
framesep=1mm,
fontsize=\footnotesize,
xleftmargin=0.2em,
]{cpp}
RUNTIME_FUNCTION(Runtime_TypedArraySortFast) {
  <..>
  size_t length = array->GetLength();
  DirectHandle<JSArrayBuffer> buffer(<...>);
  const bool copy_data = buffer->is_shared();
  // Buffer on trusted C++ heap.
  std::vector<uint8_t> offheap_copy;
  void* data_copy_ptr = nullptr;
  if (copy_data) {
    const size_t bytes = array->GetByteLength();
    if (...) {...} } else {
      // (1) Set the size of the buffer based on `bytes`
      offheap_copy.resize(bytes);
      data_copy_ptr = &offheap_copy[0];
    }
    base::Relaxed_Memcpy(data_copy_ptr, array->DataPtr(), bytes);
  }
  ctype* data = copy_data ? <...>
  // (2) Sort the array using `length` to calculate the array end
  std::sort(data, data + length);
  <...>
}
\end{minted}
\end{listing}

\stepcounter{caseStudyCtr}
\textbf{Case Study \arabic{caseStudyCtr}: Compilation Dependent Bug.}
The bug reported in \bugreport{389713719} is particularly interesting because it demonstrates that, as discussed in Section~\ref{sec:eval:abl:different_compiler_versions}, bugs may depend on how the compiler lowers the high-level \CPP code to machine code. The code that contains this bug is displayed in Listing~\ref{lst:bug_389713719_cpp}. 

\begin{listing}
\caption{Shortened code that contains bug \bugreport{389713719}. First, the constructor of \texttt{MemoryChunk} initializes the \texttt{metadata\_index\_} on-heap member variable (1). Next, this member is used to index the \texttt{metadata\_pointer\_table} table without any sanitization (2).\vspace{-1em}} 
\label{lst:bug_389713719_cpp}
\begin{minted}[
frame=lines,
framesep=1mm,
fontsize=\footnotesize,
xleftmargin=0.2em
]{cpp}
MemoryChunk::MemoryChunk(<...>): <...>,
  // (1) Write the index to `metadata_index`.
  metadata_index(MetadataTableIndex(address()))
{
  MemoryChunkMetadata** metadata_pointers = MetadataTableAddress();
  // (2) Use the metadata_index as index into
  // the `metadata_pointers` table.
  metadata_pointers[metadata_index] = metadata;
}
\end{minted}
\end{listing}

The code constructs a \verb|MemoryChunk| object inside the heap sandbox at some specific address (\verb|address()|). Since these objects contain sensitive data that must always be of integrity, the \verb|MemoryChunk| object only stores an integer index \textttWithBreak{metadata_index} in the heap sandbox (1). If sensitive data needs to be accessed, this variable is used as an index into the \textttWithBreak{metadata_pointers} that resolves to a \verb|MemoryChunkMetadata| object stored in trusted memory (2).

The issue in Listing~\ref{lst:bug_389713719_cpp} was found during a fuzzing run using a \v8 build configuration without optimization (\ie compiled with \verb|-O0|). In this configuration, an attacker may alter the value of \textttWithBreak{metadata_index} before it is used to access the lookup table, resulting in a controlled write of the value \verb|metadata|. Crucially, while the compiler must store the value \textttWithBreak{metadata_index} to the heap---since this is what the source code expresses---it is left to the compiler's discretion if it loads the value from the heap for subsequent use, or if it keeps a copy in a register or on the stack while the function is executed. No data will likely be read from the heap if the code is compiled with optimizations. Therefore, the function does not consume any attacker-controlled data. However, when the optimization level is set to \verb|-O0|, the compiler stores the index in the heap and loads it back a few instructions later, allowing an attacker to exploit this bug. Even though we may assume that unoptimized code is unlikely to be used in practice, it is important to fix such bugs. Otherwise, they may surface later due to using a different compiler or changes in the code that force the compiler to spill values to the heap that have been kept in a register before. 

\stepcounter{caseStudyCtr}
\textbf{Case Study \arabic{caseStudyCtr}: Unknown Enum Variants.}
Unlike the typical attacker models used for traditional software fuzzing, we assume an attacker has complete control of (untrusted) heap memory. This introduces a new issue: \texttt{enum} discriminants stored on the heap must be treated as having arbitrary values. We discovered multiple bugs---such as \bugreportlong{390205877}, \bugreportlong{390453039}, \bugreportlong{390568183}, and \bugreportlong{397875195}---that exploit this by constructing \texttt{enum} variants that are not defined in the source code. 
This bug class appears to be previously unknown, as an umbrella bug (\bugreportlong{390617721}) for hardening \texttt{enum} variant construction was created after we reported these issues.

Although constructing unknown \texttt{enum} variants is generally allowed, problems arise when such variants reach a sink that assumes only explicitly defined variants exist.
All bugs of this type led to undefined behavior in non-void functions that handled all explicitly defined \texttt{enum} variants but lacked a default case. As a result, these functions did not execute any switch case and reached a point where no return statement was executed, which is undefined behavior when the function has a non-void return type.

\stepcounter{caseStudyCtr}
\textbf{Case Study \arabic{caseStudyCtr}: \asan-shadowed Bugs.}
Unlike traditional fuzzing scenarios, our attacker model of the heap sandbox does not consider out-of-bounds reads as a bug (\cf Section~\ref{sec:background:attacker_model}). \asan considers such reads a bug by default, terminating the execution.
However, we found that in some instances, such as \bugreportlong{389970331} and \bugreportlong{385775375}, an out-of-bounds read is followed by an out-of-bounds \emph{write}, which constitutes a bug in our attacker model.
This demonstrates that it would be unwise to directly discard out-of-bounds reads since they can sometimes shadow more powerful primitives. Instead, we tailor \asan so that it does not terminate after encountering an out-of-bounds read.

\section{Discussion}%
\label{sec:discussion}
We now discuss directions for further research and limitations of our approach.

\textbf{Applicability to Other SFI Tools.}
Although \toolname is designed specifically for \v8, the core concept of instrumenting the fault isolation boundary can be applied to other Software Fault Isolation (SFI) implementations. For example, Firefox's RLBox~\cite{narayan2020retrofitting} isolates third-party libraries by marking data that crosses from an untrusted domain into a trusted one using annotations. These annotations define the fault isolation boundary, making it possible to insert instrumentation that a fuzzer could use to inject faults \emph{before} the data is processed by the trusted component.

\textbf{Realism of our Fault Injection Model.}
At first glance, our ability to arbitrarily modify heap sandboxed data before every intercepted load may appear overly powerful. This might seem unrealistic, as it assumes the attacker can control memory between two loads, even if execution does not return to their \javascript code in between. However, this scenario is realistic in practice: Using the \texttt{Worker} API, an attacker can spawn background threads that run concurrently with trusted engine code. These threads can modify heap memory concurrently, effectively enabling the interleaving that our fault injection model relies on (\cf~\ref{sec:background:corruption_api}). This is not just theoretical; we successfully triggered double-fetch bugs using this technique, showing that our model reflects realistic exploitation strategies (see Section~\ref{sec:eval:finding_new_vulns}).

\section{Related Work}%
\label{sec:related_work}
In the following, we discuss work related to our approach.

\textbf{Software-based Fault Isolation Schemes.}
This first SFI scheme was proposed by Wahbe~\etal~\cite{wahbe1993efficient} in 1993. Their approach uses a segment identifier (\ie some prefix bits of an address) to split the address space into multiple segments. Two of these segments, one for code, the other for stack and heap, facilitate a so-called \textit{fault domain}.
All stores and control transfer instructions that can not be proven to be within their corresponding segment are termed unsafe instructions and are instrumented.
McCamant~\etal~\cite{mccamant2005efficient} then proposed an extension, as the prior scheme can be easily bypassed on CISC architectures, such as x86, where the variable-length encoding of instructions is used. To avoid SFI checks being bypassed, McCamant~\etal propose to enforce a specific alignment and use fixed-size chunks of instructions.

Picking up this idea, Google implemented \textit{Native Client}~\cite{nacl}, which was used to execute untrusted x86 code in a browser context. In addition, bounds checks of data accesses can happen hardware-accelerated by relying on 80386 segmented memory~\cite{crawford1987programming}.
Eventually, this technique was superseded by \webassembly~\cite{WebAssemblyCoreSpecification}, a language specifically designed to compile code into a low-level format that can be executed without necessarily trusting the source of the code. This is enforced by executing \webassembly via a dedicated interpreter. %

Based on \webassembly, Narayan~\etal developed \rlbox~\cite{narayan2020retrofitting}, which compiles potentially unsafe code, such as XML parsing libraries~\cite{rlboxLibxpaImpl}, into \webassembly. Then, it runs the code in a fault-isolated environment within the same virtual address space. This system was adopted by Mozilla for Firefox to run third-party libraries that are written in C/C++. This way, Firefox cannot be compromised by bugs in these libraries. However, despite the critical importance of correctness in SFI schemes, they have received little to no rigorous testing. Our work introduces an approach to enable fuzz testing of this security boundary.

\textbf{Fault Injection.}
Prior work has already explored \emph{fault injection} in the context of fuzzing, either to test error handling code~\cite{jiang2020fifuzz, liu2021ifizz, sharma2024fuzzerr} or to mutate programs to produce slightly invalid output for the fuzz target~\cite{bars2023fuzztruction,bars2024no}.
Both use cases differ significantly from our approach, as they target a regular fuzzing scenario and focus primarily on improving error-handling code or generating more suitable program input. Our work focuses on the security boundary imposed by SFI within a single process, and we use fault injection to model the arbitrary control of an attacker within the sandbox.

\section{Conclusion}%
\label{sec:conclusion}
Software-based fault isolation represents a key technique for enforcing a security boundary that helps secure high-risk applications, such as web browsers that execute attacker-controlled code. Despite their critical importance in protecting end users, SFI schemes like \v8's heap sandbox have received surprisingly little attention from security researchers and practitioners.
In this work, we explored why existing off-the-shelf fuzzers fail to thoroughly test SFI mechanisms and introduced a novel fault injection-based method to address this gap.
Our fuzzer, called \toolname, is the first tool to be aware of this intra-process security boundary and to model an attacker's control over the sandboxed memory faithfully. Using \toolname, we uncovered \numbugs previously unknown security bugs in Google's \v8 that traditional fuzzers could not find, demonstrating its real-world applicability. We believe \toolname lays the groundwork for more rigorous testing of SFI mechanisms. We hope that future research will extend our methodology to other systems, such as Firefox's \rlbox, and contribute to building more robust and secure software isolation.

\begin{acks}

We thank the anonymous reviewers for their valuable feedback. We are also grateful to Carl Smith for his comments on an earlier draft of this work. Finally, we thank Tobias Wienand for implementing \fuzzillisbx.

This work was funded by the European Research Council (ERC) under the consolidator grant RS$^3$ (101045669) and by the Deutsche Forschungsgemeinschaft (DFG, German Research Foundation) under Germany's Excellence
Strategy (EXC 2092 CASA --- 390781972).
This material is based upon work supported by the National Science Foundation under Award No. 2232915 and by the Advanced Research Projects Agency for Health (ARPA-H) under Contract No. SP4701-23-C-0074. Any opinions, findings and conclusions or recommendations expressed in this material are those of the author(s) and do not necessarily reflect the views of the National Science Foundation or ARPA-H.
\end{acks}

\bibliographystyle{plain}
\bibliography{z_strings,z_autogenerated,z_references}
\appendix
\begin{table*}[t]
    \centering
    \small
    \caption{Different build flags used for building the \v8 engine.} 
    \label{tab:build_options}
    \begin{adjustbox}{max width=\linewidth}
 \begin{tabularx}{\linewidth}{l|X}
        \toprule
        \textbf{Flag} & \textbf{Description}\\
        \midrule
        \verb|clang_use_chrome_plugins=false| & Disables Clang plugins that are shipped with \v8 and improve its performance. Our custom Clang version may be incompatible with them. \\
        \verb|custom_toolchain| and \verb|host_toolchain| & Allows us to use our custom Clang version. \\
        \verb|is_debug=false| and \verb|dcheck_always_on=false| & Remove debug checks. Such checks would likely cause \v8 to terminate before we can observe sandbox corruption. \\
        \verb|is_asan=true| & Enables \asan as an additional bug oracle. \\
        \verb|v8_enable_sandbox=true| & Enables the heap sandbox. \\
        \verb|v8_enable_memory_corruption_api=true| & Enables the memory-corruption API required to build proof-of-concepts (PoCs) for reporting bugs. \\
        \verb|v8_fuzzilli=true| & Enables fuzzing interface used by \fuzzilli (necessary for experiments where \fuzzilli is used). \\
                \midrule
    \end{tabularx}
    \end{adjustbox}
\end{table*}

\end{document}